# A Single Photon Source based on a Long-Range Interacting Room Temperature Vapor

Felix Moumtsilis[1], Max Mäusezahl[1], Haim Nakav[1,2], Annika Belz[1], Robert Löw[1], and Tilman Pfau[1]

[1]5. Physikalisches Institut and Center for Integrated Quantum Science and Technology, Universität Stuttgart, Pfaffenwaldring 57, 70569 Stuttgart, Germany

[2]Physics of Complex Systems, Weizmann Institute of Science and AMOS, Rehovot 7610001, Israel

## 1. Context

In quantum physics, long-range interactions between neutral atoms, like dipole-dipole interactions or strong Van der Waals forces, are fundamental for understanding macroscopic quantum states. These interactions, distinct from purely local short-range counterparts, challenge existing quantum theories due to their complex dynamics, leading to new quantum states and quantum dynamics. Long-range interactions are also significant for the study of quantum entanglement and coherence, essential e.g. in quantum information processing and quantum communication. They influence the behavior of quantum gases and are instrumental in phenomena such as Bose-Einstein condensates and quantum phase transitions. These interactions underpin the development of quantum technologies, particularly in quantum computing and communication, where manipulation of entanglement and coherence over large distances is crucial.

Understanding the intricacies of these interactions provides insights into the collective behavior of quantum systems, offering potential breakthroughs in quantum simulation and computation. This includes the study of many-body systems, where long-range interactions can lead to novel quantum phases and dynamics. The field of magnetic dipolar quantum gases is reviewed in a recent publication [1].

Here we focus on the ultra-strong interaction between highly excited Rydberg atoms, which is also at the heart of the Quantum Computing platform based on neutral atoms. We show that the interaction can be so strong, that even at above room temperature quantum states of matter can be generated and used for the production of quantum states of light. In this contribution, we describe recent developments and some technical issues in our long-standing effort to investigate Rydberg interactions in hot vapor cells for quantum technological applications.

Our approach to single photon generation, employing thermal atoms in a vapor cell, contrasts with the more common use of ultra-cold atoms in quantum experiments [2]. Thermal atoms offer different dynamic properties due to their motion and dephasing. This necessitates refined techniques for controlling atomic excitation and photon emission, particularly in isolating a single highly excited state.

The utilization of thermal atoms in single photon sources is advantageous as it simplifies the experimental setup by eliminating the need for complex vacuum and cooling processes required for ultra-cold atoms. This approach also has implications for the scalability of quantum technologies, as room temperature systems can be easier to implement and maintain.

## 2. State of the art and history of the project

Our single photon source, as described by Ripka et.al. [3], utilizes thermal rubidium atoms to generate single photons. This process involves a controlled excitation of a rubidium to a Rydberg state and subsequent return to the ground state. The original idea as conceived in 2009 [4] has been under continuous development over the past decade. The initial incarnation of the single photon source required solutions to many technical issues and led to research on the interaction between glass walls and Rydberg states in a confined volume, Rydberg-Rydberg interactions, and coherent GHz Rabi cycling inside a thermal vapor.

The eventually successful design in 2018 demonstrates single photons with an correlation of $g^{(2)} \approx 0.19$, thus proving the fundamental feasibility of the system [5]. It employs a pulsed four-wave-mixing process from the $5S_{1/2}$ ground state via the $5P_{1/2}$ intermediate state to an $40S_{1/2}$ Rydberg state. This state has a Rydberg blockade radius of about 1 µm and allows only one concurrent excitation inside our tunable thickness, wedge shaped vapor cell. A thickness of only 780 nm, combined with a high numerical aperture focus of the driving lasers confines any possible Rydberg excitation to within a Rydberg blockade volume, thus creating a so-called W-state with a single excitation among all thermal atoms. A third laser drives this $40S_{1/2}$ Rydberg state to a $5P_{3/2}$ intermediate state from where it decays and emits a single photon in forward direction if the collective enhancement of the atomic vapor is sufficient. Any residual second excitation decays quickly due to the strong Rydberg-Rydberg interactions and leads to statistic emission in any direction. The whole process happens faster than any significant dephasing timescale using nanosecond pulses and GHz Rabi frequencies.

This initial proof of concept primarily suffers from the involved Rydberg transition laser wavelengths at roughly 480 nm. These wavelengths require dye amplifiers to achieve the necessary pulse powers, which are limited to repetition rates of 50 Hz and show temporal pulse jitter on the timescale of the pulse width. This hinders any ability to perform optimal control sequences, optimization, or long-term measurements. Additionally, the density of the atomic medium inside the tiny volume limits the brightness of the source. A higher number of atoms at higher temperature would increase excitation probabilities and enhance collective emission. This also requires shorter laser pulses with even higher energies to counteract the stronger dephasing at elevated temperature.

## 3. Evolution of the single photon source

After this initial demonstration and subsequent theoretical investigation for practicability, a number of concepts could improve the aforementioned shortcomings. The primary idea of each concept was to omit the unfavorable blue laser Rydberg transition by employing a different set of transitions. The excitation scheme in Fig. 1 exchanges the $5P$ intermediate states for $6P$ intermediate states, which have a much higher dipole coupling to the Rydberg state and near infrared wavelengths that can be amplified using current generation fiber technologies.

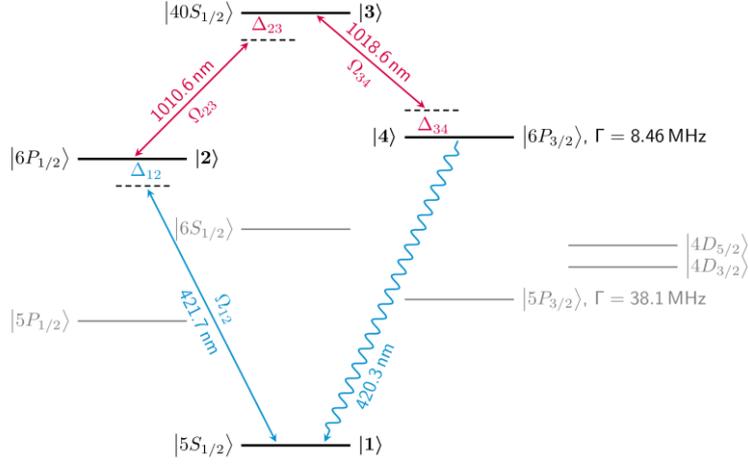

Figure 1: The evolved excitation scheme for the experiment as planned in 2018. Three lasers (lines with arrows on both ends) excite the rubidium atoms to a $6P_{3/2}$ state from where it decays into a single photon since the Rydberg blockade on the $40S_{1/2}$ state allows only a single simultaneous excitation. We enumerate the states by integers $|i\rangle$ according to the order of excitation in the pulsed laser sequence. The Rabi frequencies are labelled with $\Omega_{ij}$ and the single laser detuning from resonance is denoted by $\Delta_{ij}$ for the transition from $|i\rangle$ to $|j\rangle$.

This scheme comes at the cost of a reduced dipole coupling between the ground state and the intermediate state, which is negligible for the first laser transition from $5S_{1/2}$ to $6P_{1/2}$ and only requires a slightly stronger yet technically available laser system. One technical challenge lies in the emitted wavelength of 420.3 nm, which needs to be separated from the driving laser at 421.7 nm with at least 80 dB suppression to filter the single photons. We will discuss the adverse implications for the quantum physical emission of the emitted photons in section 6.

The construction of the required Rydberg transition laser systems as shown in Fig. 2 is a master-oscillator power-amplifier design at about 1010 nm [6]. This setup initiates with an external cavity diode laser (ECDL), and incorporates an 8 GHz bandwidth Mach-Zehnder electro-optic modulator (EOM) to create nanosecond pulses of variable length. The system further includes multiple stages of fiber amplification and acousto-optic modulators (AOMs), which are capable of producing output pulses with peak power up to 100 W, minimal picosecond-level jitter, and effectively minimized background noise (specifically, the amplification of spontaneous emission ASE). Additionally, the AOMs function as pulse-picking elements, allowing for the adjustment of the fixed 3 MHz repetition rate to a more suitable operational frequency.

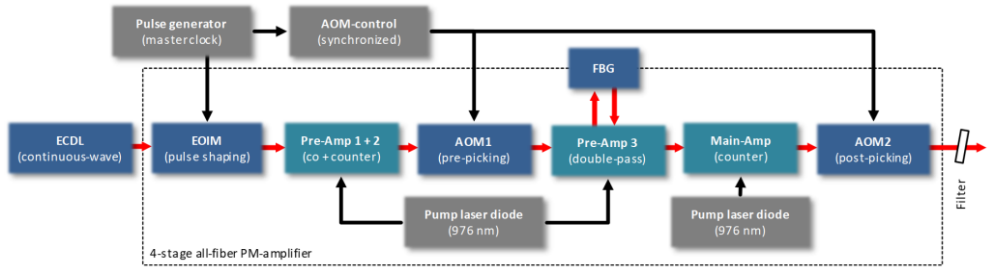

Figure 2: Schematic of the pulsed laser system used for the Rydberg transitions in the new excitation scheme (reproduced from [6]).

To increase the density of the atomic vapor beyond the levels achievable by thermal heating, we employ nanosecond light-induced atomic desorption pulses [7]. These enhance the number of participating atoms on a nanosecond timescale by up

## 4. First results using the modified excitation scheme

In order to experimentally investigate the new excitation scheme meaeuerements are taken in a thick cell length of 200 μm at a cell temperature of 190 °C. Based on the scheme in Fig. 1 the lasers are locked to resonance, pulsed with 10 ns pulse length, and propagate collinear through the cell. The experimental setup is detailed in [8]. In Fig. 3a) the Rabi frequencies of the Rydberg transitions stay constant while $\Omega_{12}$ is varied. The resulting photons at the 4th wavelength are filtered and plotted with respect to the Rabi frequency of the first transition. As a result, Rabi oscillations can be seen on the final transition in the FWM cycle, realizing a controlled excitation through the Rydberg state. Note that for times larger than 3 ns the Rabi oscillations are washed out and dissipate. However, this is expected in such a thermal ensemble due to the thermal dephasing of the atoms. Up to two Rabi cycles can be seen for higher Rabi frequencies of the first transition.

After this demonstration of Rabi cycling in the thermal system, the last transition is detuned in frequency and the resulting signal is plotted with respect to its detuning in Fig. 3b). Note that the (horizontal) trace at $\Omega_{12} = 0.5$ GHz in Fig. 3a) corresponds to the measurement at $\Delta_{12} = 0$ GHz in Fig. 3b).

We then perform this measurement for increasing temperatures and the resulting resonant trace $\Delta_{ij} = 0$ GHz is plotted with respect to the density according to the vapor pressure [9]. In Fig. 4 oscillations can be seen that increase in frequency with higher density. We attribute this to a collective enhancement of the Rabi frequencies. Thermal dephasing cannot serve as an explanation, since only the reservoir temperature is changed to reach higher densities. This results in a roughly constant velocity distribution with increasing density. Collective emission on the 4th transition promotes the de-excitation of the Rydberg W-state back to the ground state and is essential for the forward emission of photons [10].

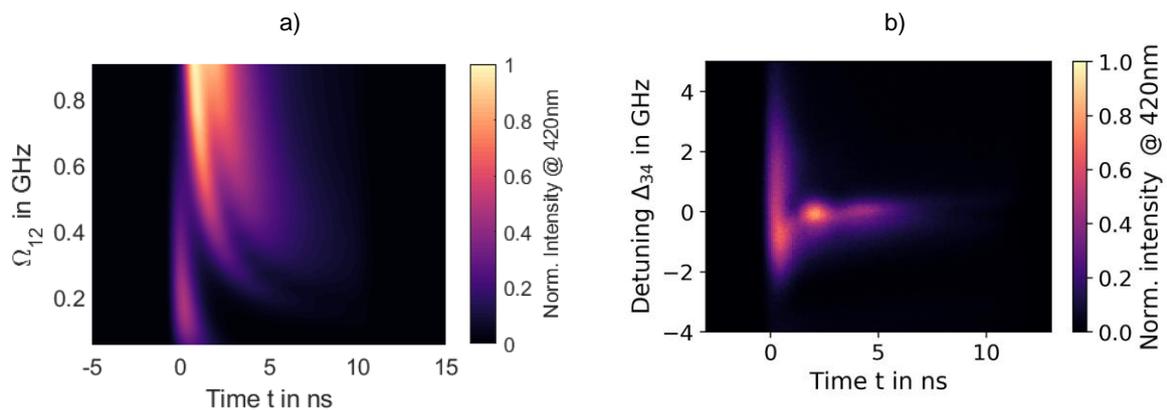

Figure 3: a) Measured signal of the emitted photons from the FWM process at a reservoir temperature of $\vartheta = 200$ °C, while the first three transitions are driven with 10 ns pulses while varying Rabi frequencies on the first transition ($\Omega_{23} = \Omega_{34} = 1.2$ GHz, all detunings $\Delta_{ij} = 0$ GHz). b) Same measurement but the third transition of the laser is frequency detuned while all other parameters stay constant (now $\Omega_{12} = 0.5$ GHz).

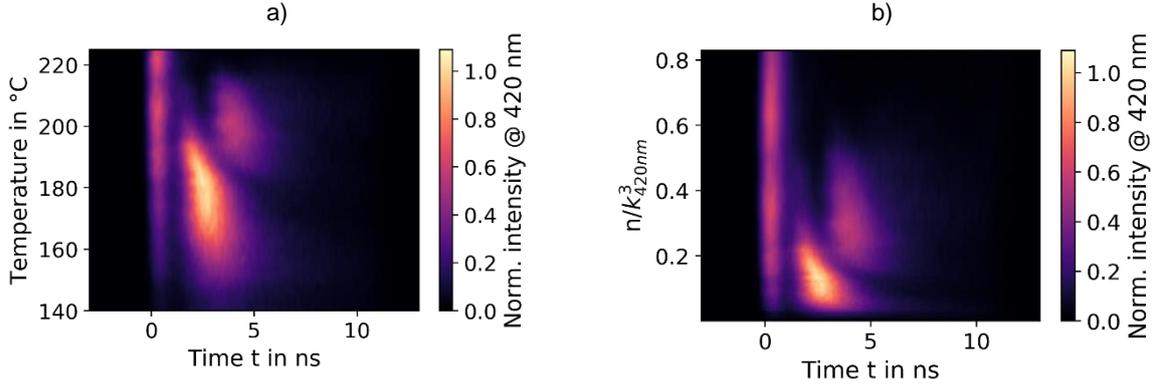

Figure 4: Temperature (a) and corresponding density (b) dependence of the Rabi oscillations of Fig. 3 at $\Delta_{ij} = 0$ GHz, $\Omega_{12} = 0.5$ GHz, and $\Omega_{23} = \Omega_{34} = 1.2$ GHz. The density is calculated from the vapor pressure according to the reservoir temperature setting using the wavenumber of the $\Omega_{12}$ transition.

## 5. Fabrication and aging of vapor cells

We produce the wedge cells at the heart of the single photon source from fused silica in a mostly traditional glass blowing process. The fabrication begins with two highly polished, approximately 1 mm thin, 50 mm long, and 25 mm wide fused silica wafers, which will form the sides of the wedge. They have a thickness precisely matched to the focusing lens to ensure a diffractions limited focal spot for the Rydberg blockade. A 20 nm atomic layer deposition (ALD) coating of alumina (sapphire) on both windows protects the fused silica from interaction with the rubidium vapor (discussed in the following paragraph). One of the flats receives an elongated chamfer of about 10 mm length and 0.5 mm depth along the short edge by lapping. We also attach a < 100 µm thick spacer to this window by adding a small glass boss looking like a small icicle and then grinding it down to a desired thickness. Clamping those parts together with a tunable force by an adjustable clamp leads to the formation of Newton's interference rings. After tuning those rings to the desired shape and thickness (which is roughly maintained during the subsequent steps), we join both windows using glass filler rod. A tube attached to the thicker end of the cell forms a reservoir in which we fill rubidium under a vacuum atmosphere.

After these wedged vapor cells are constantly kept at more than 250 °C for an extended period, they exhibit properties usually not seen in rubidium vapor experiments, e.g. in the case of reference cells. Fig. 5 shows how aging leads to an initially brown and later almost black coloration of the glass, which is no longer transparent for our lasers. The protective alumina coating does almost entirely prevent this process, even after years of operation. The coloration seems to origin from rubidium getting absorbed into the fused silica (sometimes called cell curing in other publications). At one occasion, we tried filling a cell with only a grain of rubidium (estimated to about 100 mg of solid rubidium). While this initially showed the expected behavior and increased vapor densities at 260 °C, all rubidium got consumed into the fused silica over a period of few weeks. This left the cell with a lightly brown coloration similar to the second from left specimen in Fig. 5. We confirmed that this was not due to a leak by cutting the cell open and refilling it with more rubidium. After the rightmost cell in Fig. 5 was exposed to an ambient air atmosphere the discoloration vanished over the course of several weeks. This leaves a small amount of crystalline white material behind which we estimate to be rubidium hydroxide (RbOH).

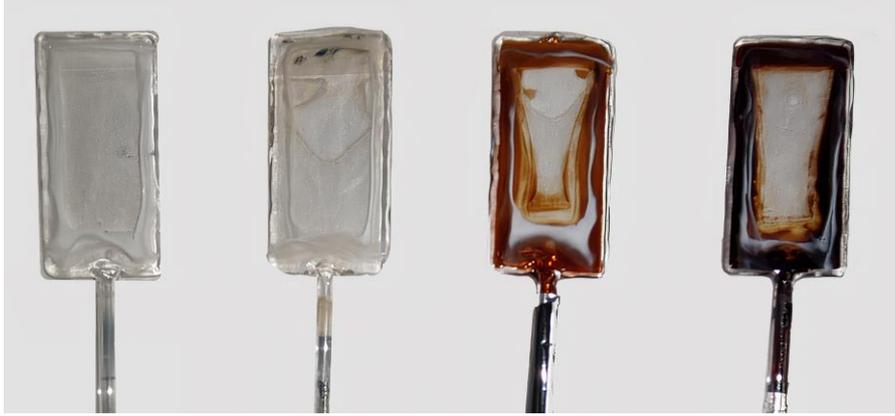

Figure 5: Pictures of the homemade wedge shaped glass vapor cells photographed in the direction, at which a laser would pass the cell in the upper transparent flat part where it has a thickness of few nm to μm. The cell on the left is new, while the cells to the right show the process of increasing aging at elevated temperature. The cell on the right was exposed to more than 250 °C for more than 2 years but stayed transparent in a central region, which is coated with a 20 nm alumina coating.

## 6. Conclusion and outlook

We demonstrated a single-photon source employing thermal rubidium atoms in a wedged microscopic glass cell. In ongoing experiments in the modified excitation scheme using a fast repetition rate fiber laser for the Rydberg excitation, we observed collectively enhanced Rabi cycling in a vapor cell.

The emission of photons on the $6P_{3/2} \rightarrow 5S_{1/2}$ transition however is more complex as compared to the $5P_{3/2} \rightarrow 5S_{1/2}$ transition. The main difference is in the branching of the spontaneous decay from the $6P_{3/2}$ state as shown in Fig. 6. As only about one fifth of all electrons will spontaneously decay from $6P_{3/2}$ to $5S_{1/2}$ the expected collectivity of the vapor is lost at a much higher rate. This means that, while the single photon character and anti-bunching are likely still present, the emission of photons does occur undirected, on a timescale longer than the dephasing time, and at multiple wavelengths as shown in Fig. 6.

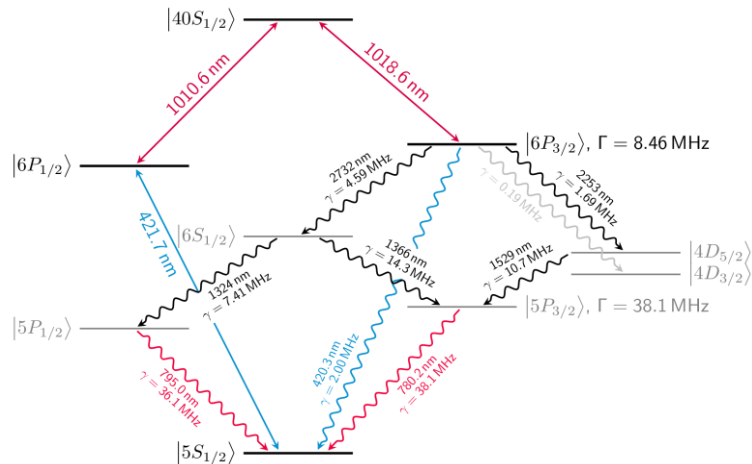

Figure 6: The same excitation scheme as shown in Fig. 1 including all possible decay channel from the final $6P_{3/2}$ state. This implies a possible reduction of the number of coherently emitted photons.

To overcome these issues and evolve the single photon source further we plan to drive the $6P_{3/2} \rightarrow 5P_{3/2}$ transition using two additional laser pulses. This will retain the technically favorable properties of the Rydberg transitions, while profiting from the branching-free, stronger, and faster spontaneous emission of the original scheme. An additional benefit of this solution is the large wavelength gap between the single photons at 780.2 nm and all driving laser wavelengths, which simplifies filtering significantly.

## 7. Acknowledgements

The authors are indebted to H. Alaeian and C. S. Adams for fruitful discussions and support. We thank F. Schreiber who performed the production of vapor cells as described in section 5.

## 8. Disclaimer

The preparation of this proceeding involved the utilization of artificial intelligence (AI) tools for initial drafting, proofreading, and refinement of the manuscript. The authors have verified the accuracy of the information presented herein to the best of their knowledge prior to publication.